\documentclass[10pt,psfig,letterpaper,twocolumn]{revtex4}
\usepackage{geometry}
\usepackage{graphics}
\usepackage{setspace}
\usepackage{natbib}
\usepackage{graphicx}
\usepackage[scaled=0.9]{helvet}

\begin{document}
\bibliographystyle{acm}

\title{\fontfamily{phv}\selectfont{\huge{\bfseries{Criteria to define a pair-ion plasma and \\  the role of electrons in nonlinear dynamics}}}}
\author{Hamid Saleem \\
 National Centre for Physics \\
Quaid-i-Azam University \\
 Islamabad, Pakistan}

\date{}
\thispagestyle{empty}
\begin{abstract}
A criterion to define a pure pair-ion (PI) plasma is presented. It
is suggested that the lighter elements (like H and He) are more
suitable to produce PI plasmas. The observation of ion acoustic wave
(IAW) in recent experiments with fullerene plasmas clearly indicates
the presence of electrons in the system. A set of two coupled
non-linear differential equations has been obtained for PI plasma
dynamics. In moving frame, it can be reduced to a form similar to
Hasegawa-Mima equation but it does not contain drift wave. {\bf
Criteria, pair-ion plasma, ion acoustic wave}
\end{abstract}

\maketitle
\section {History of the problem}

Peculiar experimental observations [1] of pair-ion (PI) fullerene $\
C^{\pm}_{60}$ plasma have invoked a great deal of interest in this
topic. It has been reported that a pure PI fullerene plasma can
support three kinds of electrostatic waves propagating parallel to
the external static magnetic field. These waves are the ion plasma
wave (IPW), the ion acoustic wave (IAW), and the third one has been
named as the intermediate frequency wave (IFW). In an earlier
experiment [2], an alkali-metal-fullerene plasma $(K^{+}, e^{-},
C^{-}_{60})$ was produced by introducing fullerenes into the
potassium plasma to realize a PI plasma. The IAW speed in Ref. [1]
has been defined as
$c_{s}=(\frac{\gamma_{i}T_{i}}{m_{i}})^{\frac{1}{2}}$ where
$m_{+}=m_{-}=m_{i}$ and $T_{+}=T_{-}=T_{i}$ have been used. Here
$\gamma_{i}$ is the ratio of ion specific heats. The subscripts plus
and minus denote the singly charged positive and negative ions,
respectively. It is important to note that here $c_{s}$k is the
frequency of the ion thermal wave and is not the IAW. This thermal
mode is similar to the sound wave in neutral fluids.

The electrostatic waves were excited in fullerene plasma externally
and following observations were noted[1]. First the IAW has
frequency larger than the theoretically calculated frequency
$v_{Ti}k$ where $v_{Ti}=(\frac{r_{i}T_{i}}{m_{i}})^{\frac{1}{2}}$ is
the ion thermal speed in our notation. Since electron density has
been assumed to be zero $(n_{eo}=0)$, therefore the ion acoustic
speed, say $c_{s}$, has not been defined as a function of electron
temperature as $c_{s}=(\frac{T_{e}}{m_{i}})^{\frac{1}{2}}$.

Second, it has been noticed that IFW has  a feature that the group
velocity is negative but the phase velocity is positive i.e. the
mode is like a backward wave. However, the IPW shows no special
features in PI plasmas.

After these observations, some theoretical investigations have shown
that the acoustic speed becomes larger in a pair-ion plasma if it is
not pure and contains significant concentration of electrons [3,4].
The IAW and some other modes have been investigated in
pair-ion-electron plasma and the linear IAW dispersion relation has
been obtained using quasi-neutrality [3]. In Ref. [4], the IAW in
PIE plasma has been discussed using kinetic approach and it has been
pointed out that the quasi-neutrality is not a good approximation
for PIE plasmas. As the number density of electrons decreases, the
electron Debye length $\lambda_{De}=(\frac{T_{e}}{4\pi
n_{e0}e^{2}})^\frac{1}{2}$ increases and the charge reparation
effects become important. Furthermore, it has been shown that the
Landau damping of IAW is reduced in the PI plasmas due to the
presence of electrons and hence this mode can be excited in the PIE
plasmas easily in the limit $1<\lambda^2_{De} k^2$. The basic
definition of plasma requires $\lambda _{De} ^{2} k ^{2} <1 $, but
the opposite limit $k \lambda _{De} ^2 k^2$ seem to be a possibel
case in PIE plasmas. The experimental set up may be very reliable to
produce PI plasmas, but the counter check is necessary to be sure
whether the produced plasma can behave as a pure PI plasma or not.
For this one needs to estimate the electron density $n_{e0}$ or the
densities of positive $n_{+0}$ and negative $n_{-0}$ ions in the
system.

One cannot say with certainty that a plasma has exactly zero
electron density. As the electron density is reduced in a system,
the electron plasma frequency $\omega_{pe}=(\frac{4\pi
n_{e0}e^2}{m_{e}})^\frac{1}{2}$ may become smaller than the
frequency of oscillations of positive ions
viz,$$\omega_{pe}<<\omega_{pi+}\eqno(1)$$

where, $\omega_{pi+}=(\frac{4\pi n_{+0} e^2}{m_{i}})^\frac{1}{2}$

We need to find out a criterion for a PI plasma. It is the
quantitative limit on the ratio of electron density to positive ion
density $(\frac{n_{eo}}{n_{+0}})$ which can decide if the role of
electrons is negligible in a plasma. Several authors [5,10] have
theoretically studied various aspects of linear and nonlinear waves
and instabilities in pure PI plasmas after the above mentioned
experimental observations. In an investigation, the behavior of the
so called backward mode (IFW) has been attributed to the nonlinear
dynamics of PI plasma [5]. Recently two more experimental research
papers have appeared in the literature on PI plasmas [11,12]. In one
of these works [11], an effort has been made to produce PI plasma
with positive and negative Hydrogen ions $(H^+ + H^-)$. On the other
hand, already a recent theoretical work [13], it has been pointed
out that a plasma of lighter ions (of low - Z materials) can behave
as a pure PI plasma with relative larger electron density $n_{e0}$
compared to plasmas of heavier ions like fullerenes. Therefore, it
has been suggested that it is more suitable to try to produce PI
plasmas using low-Z materials if other physical conditions like the
electron attachment cross-section (to produce negative ions),
ionization and recombination rates can be controlled. Some linear
and nonlinear waves of PI and PIE plasmas are also discussed in this
work.

\section {IAW in PIE Plasmas}

Here we derive the linear dispersion relation of ion acoustic wave
(IAW) in pair-ion-electron (PIE) plasma for the simplest case,
$T_{+}=T_{-}=T _{i}<<T _{e}$ using fluid equations. The wave is
assumed to propagate along the external magnetic field to compare
the theoretical result with the experimental observation [1].

Let the constant external magnetic field be along z-axis, i.e. ${\bf
B_{0}}=B_{0}{\bf z\hat{}}$ and the wave be propagating along the
field lines with wave vector ${\bf k}\vec{}=k{\bf z\hat{}}$. The
equations of motion for singly charged positive and negative ions
become, respectively,
$$\partial_{t}v_{+z}\simeq
-\frac{e}{m_{i}}\partial_{z}\varphi\eqno(2)$$

and $$\partial_{t}v_{-z}\simeq
+\frac{e}{m_{i}}\partial_{z}\varphi\eqno(3)$$

The continuity equations yield, $$\partial_{t}(n_{+}-n_{-}) + n_{+0}
\partial_{z} v_{+z} - n_{-0} \partial_{z} v_{-z}=0\eqno(4)$$

Assuming electrons to follow the Boltzmann density distribution in
electrostatic field $E=-\Delta\varphi$ as, $$n_{e}\simeq n_{e0}
e^\frac{e\varphi}{T_{e}}\simeq n_{e0}
(1+\frac{e\varphi}{T_{e}})\eqno(5)$$

the Poisson equation gives, $$(n_{+}-n_{-})\simeq -\frac{\\
\nabla^2\varphi}{4\pi e}+n_{e0} \frac{e\varphi}{T_{e}}\eqno(6)$$

Using the above set of equations, we can obtain the linear
dispersion relation as, $$\omega^2 _{s}=N_{0} \frac{c^2
_{s}k^2}{1+\lambda^2 _{De}k^2}\eqno(7)$$

where $N_{0}=(\frac{n_{+0}+n_{-0}}{n_{e0}})$. This is the same as
Eq. [5] of Ref.[4]. Note that $c_{s}k<\omega_{s}$ when
$n_{e0}<n_{+o}$ because $1<N_{0}$. The IAW can be excited easily in
PIE plasma because the Landau damping rate decreases in the limit
$1<\lambda _{De} ^{2} k^{2}$[4].

\section {Criteria for Pair-Ion Plasma}

To decide whether the produced plasma can be called a pure PI plasma
or not, one needs to estimate the ratio $\frac{n_{e0}}{n_{+i}}$. If
the electrons pressure is not unusually high due to some external
heating mechanism, then the situation
$\omega_{pe}^{2}<<\omega_{pi+}^{2}$ implies that their role in
plasma dynamics can be neglected. Such a system can behave as a pure
PI plasma. Let us consider the plasma wave dispersion relation using
kinetic approach written as,
\\$$1+\sum_{j}\frac{1}{k^{2}\lambda_{Dj}^{2}}\{1+\iota\sqrt{\pi}Z_{j}W(Z_{j})\}=0\eqno(8)$$
\\where $\lambda_{Dj}^{2}=\frac{T_{j}}{4\pi n_{0j}e^2}$;
$Z_{j}=\frac{\omega}{\sqrt{2}k\nu_{Tj}}$;
$v_{Tj}=(\frac{T_{j}}{m_{j}})^\frac{1}{2}$ and $W(Z_{j})$ is the
plasma dispersion function for the jth species $(j=\pm,e)$. In
equilibrium, the quasineutrality demands $n_{0e}+n_{-0}=n_{+0}$
where both the positive and negative ions are assumed to be singly
charged.This is the case of Ref. [1] as well. For the time being,
let us assume $T_{i}<T{e}$ and $\ v_{Te}<<\frac{\omega}{k}$ such
that $1<<|Z_{e}|<<|Z_{i}|$ which allows us to use an asymptotic
expansion of $W(Z_{j})$ to study the system analytically. The case
of IAW with $v_{Ti}<<\frac{\omega}{k}<<v_{Te}$ has been discussed in
Ref. [4] in detail.

Assuming $m_{+}=m_{-}=m_{i}$, Eq. [8] can be expressed as,
$$\omega^{2}-P_{n}^{0}\omega_{pi+}^{2}-\frac{3}{\omega^{2}}\{k^{2}\ v_{Te}^2
\omega_{pe}^2+k^{2}v_{Ti}^{2}-\omega_{pi-}^{2}+k^{2}v_{Ti+}^{2}+\omega_{pi+}^{2}\}$$
$$+\iota\sqrt{\pi}\omega^{2}\left\{\frac{Z_{e}}{k^{2}\lambda_{De}^{2}}e^{-Z_{e}^{2}}+
\frac{Z_{i-}}{k^{2}\lambda_{Di-}^{2}}e^{-Z_{i}^{2}}+\frac{Z_{i+}}{k^{2}\lambda_{Di+}^{2}}e^{-Z_{i+}^{2}}\right\}$$
$$=0\eqno(9)$$
\\where
$P_{n}^{0}=(1+\frac{n_{-0}}{n_{+0}}+\frac{n_{e0}}{n_{+0}}\frac{m_{i}}{m_{e}})$.
For $\omega=\omega_{r}-i\gamma$, the real and imaginary parts of Eq.
[9] become, respectively, $$\omega_{r}^{2}\simeq
P_{n}^{0}\omega_{pi\pm}^{2}+\frac{3}{P_{n}^{0}}[\nu_{Ti\pm}^{2}k^{2}+(\frac{\omega_{pi-}^{2}}{\omega_{pi+}^{2}}k^{2}\lambda_{Di-}^{2})\omega_{pi-}^{2}$$
$$+(\frac{\omega_{pe}^{2}}{\omega_{pi+}^{2}}k^{2}\lambda_{De}^{2})\omega_{pe}^{2}]\eqno(10)$$

and
$$\gamma\simeq\sqrt{\frac{\pi}{4}}[\frac{Z_{i+}}{k^{2}\lambda_{Di+}^{2}}\exp\{-Z_{i+}^{2}\}+\frac{Z_{i-}}{k^{2}\lambda_{Di-}^{2}}\exp\{-Z_{i-}^{2}\}$$
$$+\frac{Z_{e}}{k^{2}\lambda_{De}^{2}}\exp\{-Z_{e}^{2}\}]\omega_{r}\eqno(11)$$

The important point to note is that in the limit
$\omega_{pe}<\omega_{pi+}$ and $v_{Te}k<\omega$ the electron plasma
wave turns into the ion plasma wave. Furthermore in the limit
$(\frac{n_{0e}}{n_{i+}})\rightarrow0$, we need not to use the
perturbed electron density $n_{e1}$ in the Poisson equation.

The comparison between the two terms of Eq. [10], i.e.,
$k^{2}v_{Te}^{2}\omega_{pe}^{2}$ and
$k^{2}v_{Ti+}^{2}\omega_{pi+}^{2}$can be very important to decide if
the plasma can be called a pure (PI) system. Note that if
$T_{i}=T_{i+}=T_{i-}$ and $m_{i+}=m_{i-}=m_{i}$, then
$k^{2}v_{Te}^{2}\omega_{pe}^{2}<<k^{2}v_{Ti}^{2}\omega_{pi+}^{2}$
provided that
$$\frac{n_{0e}}{n_{+0}}<<\frac{T_{i}}{T_{e}}\left(\frac{m_{e}}{m_{i}}\right)^{2}\eqno(12)$$

In this case the electron contribution to the plasma dynamics can be
neglected only for a very small value of the ratio
$\frac{n_{0e}}{n_{i+}}$. Then the plasma can be called a pure (PI)
plasma.

In the case of fullerene plasma, $m_{i}=720m_{p}$ (where $m_{p}$ is
mass of the proton) and $\frac{m_{e}}{m_{p}}\sim\frac{1}{1836}$.
Therefore we have $\frac{m_{e}}{m_{i}}=7.56\times10^{-7}$. If
$T_{i}<T_{e}$ is assumed, then the fullerene plasma discussed in
Ref. [1] can be called a pure pair-ion plasma only if the following
limit holds:
$$\frac{n_{0e}}{n_{i}}<<(7.56\times10^{-7})^{2}\frac{T_{i}}{T_{e}}\eqno(13)$$

The plasma density in Ref. [1] is $n_{i}\sim 10^7 cm^{-3}$. It means
that this system can become a pure (PI) plasma only if there is no
electron in the system which is very unlikely physically.

Fortunately, we have a better condition than (12) to call the plasma
a pure (PI) plasma. Our main requirement is the limit
$\omega_{pe}^{2}<<\omega_{\pm}^{2}$, which replaces the relation
(12) by a new limit as follows:
$$\frac{n_{0e}}{n_{\pm}^{0}}<<\alpha\frac{m_{e}}{m_{i}}\eqno(14)$$

where $\alpha$ must satisfy the condition $\alpha<<1$.
Correspondingly the condition on thermal correction term is,
$$k^{2}v_{Te}^{2}\frac{\omega_{pe}^{2}}{\omega_{pi+}^{2}}\leq\omega_{pi+}^{2}\eqno(15)$$
Since $\frac{\omega_{pe}^{2}}{\omega_{pi+}^{2}}\simeq\alpha$
therefore (15) suggests the following maximum limit on the value of
$\alpha$ to call a plasma as PI plasma:
$$c_s k<<\alpha k \nu_{Te}<<\omega_{pi+}\eqno(16)$$

If we choose $m_e/m_i<<\alpha<<1$ the condition (16) is satisfied
and it is then in agreement with the fact that the ion acoustic wave
should not appear in the pure (PI) plasma. However, a smaller value
of $\alpha$ is preferable. If $n_{eo}$ is so small that $\alpha$ is
almost zero, then we will have,
$$\alpha v_{Te}^{2} k^2<<c_{s}^{2}k^2<<<\omega_{pi+}^{2}\eqno(17)$$

In  this case the IAW remains almost non-existent. That is $n_{eo}$
is too small and electron pressure does not  contribute to plasma
dynamics. The only normal mode of the system with
$\mathbf{k}||\mathbf{B}_0$ is the ion plasma wave which may have a
negligible contribution from the small number of hot electrons. For
the case of Helium (He) plasma $m_e/m_p\simeq10^{-4}$ and if
$T_i<T_e$ is assumed, then it will become almost a pure (PI) plasma
if $n_{0e}/n_{i}<<10^{-5}$ holds.  Therefore  we conclude that it is
more suitable to try to produce (PI) plasma of lighter atoms (or
molecules) if other physical conditions like  the
ionization/recombination rates, and the electron attachment cross
section  can be controlled. Therefore it is suggested that the
Hydrogen and Helium systems can be very useful to achieve a pure PI
plasma.

\section {Vortices in PIE Plasmas}

If $T_{i}\neq0$, then the perpendicular drift velocities for ions
can be written as,
$$\textbf{v}_{j\perp}=\frac{c}{B_{0}}\textbf{E}_{\perp}\times\textbf{z}-
\frac{\nabla_{pj}\times\textbf{z}}{\Omega_{j}m_{j}n_{j}}-
\frac{1}{\Omega_{i}}(\nabla_{i}+\textbf{v}_{j}.\nabla)\textbf{v}_{j}\times\textbf{z}$$
$$=\textbf{v}_{E}+\textbf{v}_{Dj}+\textbf{v}_{pj}\eqno(18)$$ Here
$j=\pm$ and $\Omega_{j}=\frac{q_{j}B_{0}}{m_{j}c}$. For electrons,
we have
$$\textbf{v}_{e\perp}=\frac{c}{B_{0}}\textbf{E}_{\perp}\times\textbf{z}-
\frac{\nabla_{pj}\times\textbf{z}}{\Omega_{e}m_{e}n_{e}}=\textbf{v}_{E}+\textbf{v}_{De}\eqno(19)$$
where $|\partial_{t}|<<\Omega_{e}=\frac{eB_{0}}{m_{e}c}$ has been
used. The continuity equations of the ions yield,
\\\\\\
$$\partial_{t}(n_{+}-n_{-})+\frac{c}{B_{0}}\nabla n_{e0}.(\textbf{Z}\times\nabla_{\perp\phi})
-\frac{c}{B_{0}\Omega_{i}}(n_{+0}+n_{-0})$$
$$\times(\partial_{t}+\textbf{v}_{E}.\nabla)\nabla^{2}\phi
=n_{-0}\partial_{z}\ v_{z-}-n_{+0}\partial_{i}\ v_{z+}\eqno(20)$$
and the parallel equation of motion becomes,
$$(\partial_{t}+\textbf{v}_{E}.\nabla)=n_{-0}\partial_{z}\
v_{z-}-n_{+0}\partial_{i}\nu_{z+}$$
$$=\frac{e}{m_{i}}(n_{+0}+n_{-0})\partial_{z}\phi\eqno(21)$$ Assuming
Boltzmann density distribution for electrons $n_{e}\sim
n_{e0}e^{\frac{e\phi}{T_{e}}}$ and using the Poisson equation.
$$\nabla.\textbf(E)=4\pi e(n_{+}-n_{-}-n_{e})\eqno(22)$$
the nonlinear Eqs. [19] and [20] can be written, respectively, as:
$$\partial_{t}\{-\lambda_{De}^{2}\nabla^{2}\Phi\}+D_{e} \kappa_{ne}.(\textbf{z}\times\nabla_{\perp}\Phi)-N_{0}\rho_{s}^{2}$$
$$\times(\partial_{t}+D_{e}\textbf{z}\times\nabla_{\perp}\Phi.\nabla)\nabla_{\perp}^{2}\Phi=\partial_{z}V\eqno(23)$$
and
$$(\partial_{t}+D_{e}\textbf{z}\times\nabla_{\bot}\Phi.\nabla)V=c_{s}^{2}N_{0}\partial_{z}\Phi\eqno(24)$$
where $V=\frac{(n_{-0}v_{-z}-n_{+0}v_{+z})}{n_{e0}}$,
$N_{0}=\frac{(n_{+0}+n_{-0})}{n_{e0}}=\frac{e\phi}{T_{e}}$,
$D_{e}=\frac{eB_{0}}{cT_{e}}$,
$\rho_{s}^{2}=\frac{c_{s}^{2}}{\Omega_{i}^{2}}$ and
$c_{s}=(\frac{T_{e}}{m_{i}})^{\frac{1}{2}}$.

Equation [23] is the Hasegawa-Mima (HM) equation for (PIE) plasma if
the RHS is ignored (for $\ v_{iz}\rightarrow0$) and
$\lambda_{De}^{2}k^{2} << 1$ is assumed.

These equations give the coupled linear dispersion relation of drift
wave and IAW in (PIE) plasmas as,
$$G_{0}\omega^{2}-\omega_{e}^{*}-N_{0}c_{s}^{2}k_{0} ^{2}=0\eqno(25)$$
where
$G_{0}=(1+\lambda_{De}^{2}k^{2}+N_{0}\rho_{s}^{2}k_{\perp}^{2})$ and
$\omega_{e}^{*}=\textbf{v}_{0}^{*}.\textbf{k}$,

If $N_{0}c_{s}^{2}k_{\perp}^{2}<<\omega_{e}^{*}$ holds, then we
obtain only the drift wave dispersion relation as,
$$\omega=\frac{\omega_{e}^{*}}{(1+\lambda_{De}^{2}k^{2}+N_{0}\rho_{s}^{2}k_{\perp}^{2})}\eqno(26)$$
In (PIE) plasma the quasi-neutrality can break down for IAW in the
limit $1<<\lambda_{De}^{2}k^{2}$ because $n_{e0}$ can be very small.
On the other hand the inequality $\lambda_{De}<\rho_{s}$ always
holds. Since $\rho_{s}^{2}k^{2}<1$ in the fluid model, therefore
$\lambda_{De}^{2}k^{2}$ should not be much larger than 1. This means
in magnetized plasmas, the IAW cannot have wavelengths shorter than
$\lambda_{De}$ within fluid theory framework because of the limit,
$$\lambda_{De}^{2}k^{2}<\rho_{s}^{2}k_{\perp}^{2}<1\eqno(27)$$

It is important to note that as $n_{e0}$ decreases, $N_{0}$
increases to have
$\lambda_{De}^{2}k^{2}<<N_{0}\rho_{s}^{2}k_{\perp}^{2}$ and hence
Eqs. [23] and [24] give the pair plasma convective cell (PPCC) mode,
$$\omega^{2}=\frac{k_{s}^{2}}{k_{\perp}^{2}}\Omega_{i}^{2}$$

As $n_{e0}$ decreases, the IAW converts into the PPCC mode. In
between these two limits, the electron drift wave couples with IAW
and PPCC.

\section {Nonlinear Equations for PI Plasma Dynamics}

The set of nonlinear Eqs. [23] and [24] can be transformed into HM
equation in a moving frame which admits monopolar and dipolar vortex
solutions [13].The nonlinear dynamics of (PI) plasmas are described
by Eqs. [23] and [24] in the limit $1<<N_{0}$ and they reduce,
respectively, to the following equations:
$$(\partial_{t}+D_{i}\textbf{z}\times\nabla_{1}\Phi.\nabla)\nabla_{\perp}^{2}\Phi=-\frac{1}{2\rho_{i}^{2}}\partial_{z}V\eqno(28)$$
and
$$(\partial_{t}+D_{i}\textbf{z}\times\nabla_{\perp}\Phi.\nabla)V=2v_{Ti}^{2}\partial_{z}\Phi\eqno(29)$$
where $\Phi=\frac{e\phi}{T_{i}}$, $D_{i}=\frac{cT_{i}}{eB_{0}}$,
$\rho_{i}=\frac{\nu_{Ti}}{\Omega_{i}}$ and $V=(v_{z-}-v_{z+})$\\

In the stationary $(\eta,x)$ frame, the coupled Eqs. [28] and [29]
can be written as,
$$C_{1}d_{\eta}\nabla_{\perp}^{2}\Phi+V_{0}d_{\eta}\Phi+\{\nabla_{\perp}^{2}\Phi,\Phi\}=0\eqno(30)$$
where $C_{1}=\frac{-u}{D_{i}}$, $V_{0}={\mu
L_{0}}{2\rho_{i}^{2}D_{i}}$,
$L_{0}=\left(C_{0}-\frac{2\mu\nu_{Ti}^{2}}{u}\right)$ and $C_{0}$ is
an arbitrary constant. The important point to note is that the form
of Eq. (28) is similar to Hasegawa-Mima equation but the physics of
the equation is completely different. The set of nonlinear Eqs. [28,
29] is valid as well for electron-position plasmas in the classical
limit. But these equations do not contain the drift wave and the ion
acoustic mode. They describe the nonlinear dynamics of (PI) plasmas
in the quasi-neutrality approximation.

\section {PI Plasma For Fusion}

The drift waves are the fundamental source of instabilities in
tokamak fusion plasmas. Several kinds of reactive and dissipative
drift instabilities appear in Tokamak plasmas. The presence of
electrons in laser-fusion is also problematic for laser absorption,
heat conductions and uniform compression. The PI-plasmas can be very
suitable fuel for fusion, in principle because drift waves cannot
exits in such plasmas. However, it does not seem easy, at least at
present times, first to produce PI plasma of Hydrogen or Helium at
high densities and high temperatures for fusion. Second the
confinement for a longer time can also be a problem because of
recombination and production of electrons and neutrals as a result
of collisions between positive and negative ions as well as with
neutrals.

\section {Summary}

A criteria to define a pure pair-ion (PI) plasma has been discussed.
The condition (14) must be satisfied along with (16) to call a
plasma as a pure PI plasma. But a very small value of $\alpha$ such
that $\alpha<<\frac{m_{e}}{m_{i}}$ is preferable. It is suggested
that the lighter elements are preferable to produce pure PI plasma
if electron attachment cross-section and recombination rate can be
controlled for desirable results. In the plasma of lighter elements,
the condition (14) can be satisfied even for relatively larger
values of $\alpha$.

It has also been stressed that the observation of ion acoustic wave
(IAW) in the experiment, itself is an indication of the existence of
significant concentration of electrons in the produced pair-ion
fullerene plasma and hence it cannot behave as a pure PI plasma.
Furthermore, the frequency of IAW increases in a pair-ion plasma in
the presence of electrons because we have $1<<N_{0}$.

A set of two coupled nonlinear differential equations has also been
obtained for the pure PI plasma. In a moving frame, these equations
can be reduced to a single equation similar to Hasegawa-Mima
equation but it does not contain drift wave.

{}

\end{document}